\documentclass[aps,prl,twocolumn,amsmath,amssymb,floatfix,superscriptaddress]{revtex4-1}
\usepackage{tabularx}
\usepackage{bm}
\usepackage{euscript}
\usepackage{epsfig,psfrag,subfigure}
\usepackage{graphicx}
\usepackage{color}
\usepackage{amsfonts}
\usepackage{exscale}
\usepackage{amsbsy}
\usepackage{multirow}
\usepackage[normalem]{ulem}
\usepackage[toc,page,title,titletoc,header]{appendix}
\usepackage[colorlinks,linkcolor=blue,citecolor=blue,anchorcolor=blue]{hyperref}

\begin{document}

\title{Ground state phase diagram and superconductivity of the doped Hubbard model on six-leg square cylinders}

\author{Yi-Fan Jiang}
\affiliation{School of Physical Science and Technology, ShanghaiTech University, Shanghai 201210, China}

\author{Thomas P. Devereaux}
\affiliation{Stanford Institute for Materials and Energy Sciences, SLAC National Accelerator Laboratory and Stanford University, Menlo Park, CA 94025, USA}
\affiliation{Department of Materials Science and Engineering, Stanford University, Stanford, CA 94305, USA}

\author{Hong-Chen Jiang}
\email{hcjiang@stanford.edu}
\affiliation{Stanford Institute for Materials and Energy Sciences, SLAC National Accelerator Laboratory and Stanford University, Menlo Park, CA 94025, USA}

\begin{abstract}
We have studied the ground state properties of Hubbard model on long six-leg square cylinders with doped hole concentration per site $5.55\% \leq \delta\leq 12.5\%$ using density-matrix renormalization group. By keeping a large number of states for long system sizes, we find that the nature of the ground state is remarkably sensitive to the presence of next-nearest-neighbor electron hopping $t'$. In the positive $t'$ side, we find a robust $d$-wave superconducting (SC) phase characterized by coexisting quasi-long-range SC and charge density wave (CDW) correlations. Without $t'$ the ground state forms an insulating stripe phase with long-range CDW order but short-range spin-spin and SC correlations. In stark contrast to four-leg cylinders, our results show that the lightly doped Hubbard model on six-leg cylinders remains insulating in the negative $t'$ side where the SC correlations decay exponentially with short correlation lengths. In the larger negative $t'$ side, the doped holes form a novel holon Wigner crystal with one doped hole per emergent unit cell and short-range spin-spin correlations.
\end{abstract}
\date{\today}

\maketitle

%\section{Introduction}
The Hubbard model plays a paradigmatic role in the theory of strongly correlated many-body systems \cite{Dagotto1994,Scalapino2012,Zhang2014,Keimer2015,Arovas2022,Qin2022}. It is widely believed that this seemingly simple model could exhibit strikingly rich quantum phases including various forms of anti-ferromagnetism, charge density waves, and unconventional superconductivity. However, despite tremendous efforts devoted over more than half a century, various basic properties of the actual 
phases within Hubbard models still remain controversial. This is partially due to the insufficiency of controlled analytical approaches for strongly correlated systems and the prevalence of many low-energy competing orders. However, with significant numerical method developments in recent years, many progresses have been made to understand properties of various quantum phases resulting from moderate interaction strengths \cite{Arovas2022,Qin2022, White1997, Lichtenstein2000, Sorella2002, Capone2006,Tocchio2008, Yang2009, Corboz2011, Gull2012, Misawa2014, Corboz2014, LeBlanc2015, Tocchio2016, Huang2017, Ehlers2017, White1999, Scalapino2012W, Dodaro2017,Zheng2017,Qin2020, Huang2018, Ido2018, Ponsioen2019, Tocchio2019, Sorella2021}. 
From controlled numerical treatments, especially using density-matrix renormalization group (DMRG) \cite{White1999,Scalapino2012W,Dodaro2017,Zheng2017,Qin2020,Arovas2022,Qin2022}, there is a growing consensus that unidirectional charge-density-wave (CDW) (i.e. “stripe”) order \cite{Zaanen1989,Zaanen1998,Zaanen2001} rather than superconductivity arises in lightly doped ``pure" Hubbard models having only a nearest-neighbor (NN) electron hopping $t$ and intermediate-value Coulomb repulsion $U$.

%==Fig.1 Phase diagram==
\begin{figure}[bt]
\centering
\includegraphics[width=\linewidth]{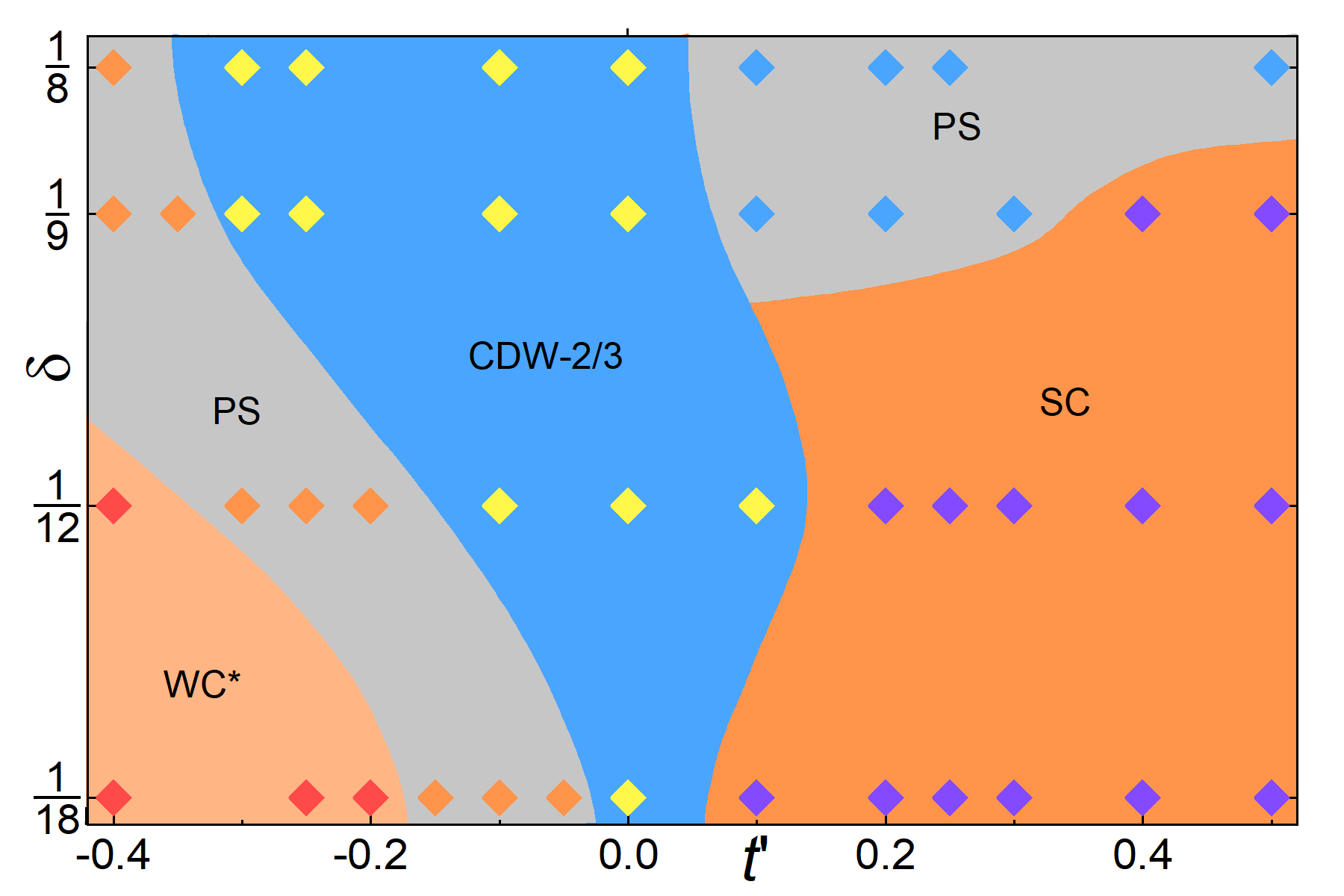}
\caption{(Color online) Ground state phase diagram of the Hubbard model in Eq.(\ref{Eq:Ham}) as a function of $t'$ and  hole doping concentration $\delta$ at $U$=12 where the solid diamonds are data points. Here the $d$-wave superconducting phase is denoted by SC, holon Wigner crystal is denoted by WC$^*$, ``two-third-filled" charge stripe phase is denoted by CDW-2/3 and phase separation is denoted by PS.}
\label{Fig:PhaseDiagram}
\end{figure}

However, recent numerical studies have shown that the balance between superconductivity and other forms of order, such as CDW order, can be sensitively tipped by the inclusion of next-nearest-neighbor (NNN) electron hopping $t'$. For instance, superconducting (SC) correlations can be significantly enhanced by adjusting $t'$, where recent DMRG studies have shown that a robust quasi-long-range superconductivity can be achieved in the doped Hubbard model on four-leg square cylinders for both positive and negative $t'$ \cite{Jiang2018tJ,Jiang2019Hub,Jiang2020Hub,Peng2022}. Contrary to the normal $d$-wave pairing symmetry observed for electron doping with positive $t'$, a plaquette-type $d$-wave symmetry is observed in the hole-doping with negative $t'$, which is unique to four-leg square cylinders \cite{Dodaro2017,Chung2020}, inconsistent with what might be expected for 2D systems. 
Similarly, SC correlations can also be significantly enhanced by $t'$ in doped $t$-$J$ models, i.e., the strong coupling limit of the Hubbard model, on systems wider than four-leg cylinders, compared with the "pure" Hubbard model without $t'$. However, this enhancement is only observed for electron-doping with positive $t'$ \cite{Gong2021,Jiang2021White,Jiang2021tJ,Jiang2022White,Jiang2023,Lu2023Sign}. This is surprising in the context of high temperature superconducting cuprates given that the band dispersions of hole-doped cuprates require negative $t'$. Whether the SC correlations can be notably enhanced in the electron-doped Hubbard model with positive $t'$, and whether robust superconductivity can also be achieved in the hole-doped Hubbard model on systems wider than four-leg cylinders, especially for intermediate interactions, where $U$ is comparable to the bandwidth of the system, has remained elusive.

{\bf Principal results: }%
In this paper, we present extensive DMRG studies of the $t$-$t'$-$U$ Hubbard model at hole doping concentration of $\delta=1/18 - 1/8$ and for $-0.4\leq t'/t\leq 0.5$, carried out on six-leg square cylinders with periodic and open boundary conditions in the short and long directions, respectively. Our main results are summarized in the ground state phase diagram of the Hubbard model in Fig.\ref{Fig:PhaseDiagram}, which is surprisingly sensitive to both $t'$ and $\delta$. The blue region around $t'\sim 0$ is identified as an insulating phase with a unidirectional ``2/3-filled" charge stripes and mutually commensurate spin stripes, but short-range SC correlations. This charge stripe phase is similar to the one reported in previous DMRG studies of the ``pure" Hubbard model with $t'=0$ \cite{Jiang2019Hub,Jiang2020Hub,Qin2020}. In the presence of positive $t'$, a robust $d$-wave SC phase, similar to the one observed in the Hubbard model on four-leg square cylinders \cite{Jiang2020Hub} and the closely related $t$-$J$ model on wider systems \cite{Gong2021,Jiang2021White,Jiang2021tJ,Jiang2022White,Jiang2023,Lu2023Sign}, is found. This phase is characterized by coexisting quasi-long-range SC and CDW orders, but short-range spin-spin correlations.

However, in stark contrast to four-leg cylinders, we find that the doped Hubbard model on six-leg cylinders in the hole-doped case with negative $t'$ appears insulating with strong CDW order. Other correlations, including both SC and spin-spin correlations, are short-ranged with fairly short correlation lengths. Surprisingly, in addition to the usual charge stripe order, a distinct CDW phase appears at the lower-left corner of the phase diagram with relatively larger negative $t'$ and lower hole doping concentrations. This CDW phase, which we refer to as holon Wigner crystal (WC$^*$) where the holon carries the charge of an electron but without its spin\cite{Rokhsar1988}, is similar with the one observed in the doped spin liquid on the Kagome lattice \cite{Jiang2017,Peng2021}. In this phase, the doped holes form a long-ranged CDW ordered state with one doped hole per emergent unit cell, while both SC and spin-spin correlations are short-ranged. Therefore, this is a crystal of spinless holons instead of holes.

%Properties of d-wave SC phase
\begin{figure}[bt]
\centering
\includegraphics[width=\linewidth]{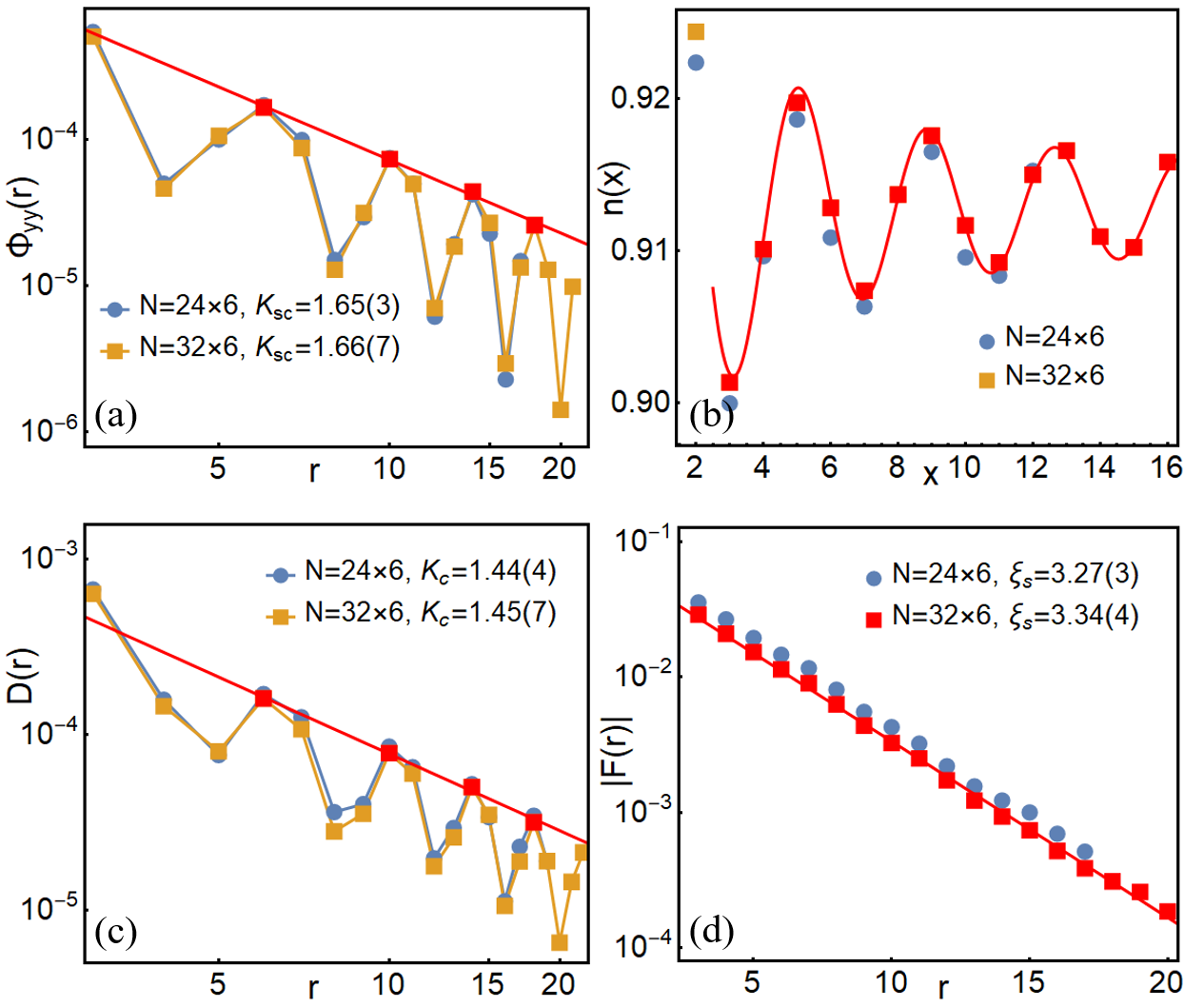}
\caption{(Color online) Characterization of the $d$-wave SC phase: (a) Superconducting correlation $\Phi_{yy}(r)$ plotted on double-logarithmic scales where the solid line denotes power-law fit $\Phi(r)\sim r^{-K_{sc}}$. 
(b) Charge density profiles $n(x)$ fitted by the Friedel oscillation (solid line) using Eq.(\ref{Eq:Kc}). (c) Charge density-density correlation $D(r)$ on double-logarithmic scales where the solid line denotes a power-law fit $D(r)\sim r^{-K_{c}}$. (d) Spin-spin correlation $F(r)$ on a semi-logarithmic scale where the solid line denotes an exponential fit $|F(r)|\sim e^{-r/\xi_s}$. Here $t'=0.5$, $\delta=1/12$, and data points in red are used for the fit. 
}
\label{Fig:phase_SC}
\end{figure}

{\bf Model and Method: }%
We employ DMRG \cite{White1992, Schollwock2005} to investigate the ground state properties of the lightly doped Hubbard model on square lattice defined by the Hamiltonian%
\begin{eqnarray}\label{Eq:Ham}
H =- \sum_{ij\sigma} t_{ij} \left(\hat{c}^\dagger_{i\sigma} \hat{c}_{j\sigma} + h.c.\right)+ U\sum_i \hat{n}_{i\uparrow}\hat{n}_{i\downarrow}. 
\end{eqnarray}
Here $\hat{c}^\dagger_{i\sigma}$ ($\hat{c}_{i\sigma}$) is the electron creation (annihilation) operator on site $i=(x_i,y_i)$ with spin $\sigma$, $\hat{n}_{i\sigma}=\hat{c}^\dagger_{i\sigma}\hat{c}_{i\sigma}$ and $\hat{n}_i=\sum_{\sigma}n_{i\sigma}$ are the electron number operators. The electron hopping amplitude $t_{ij}$ is equal to $t$ if $i$ and $j$ are NN and equal to $t^\prime$ if $i$ and $j$ are NNN. $U$ is the on-site Coulomb repulsion. We take the lattice geometry to be cylindrical and a lattice spacing of unity. The boundary condition of the cylinders is periodic in the $\hat{y}=(0,1)$ direction while open in the $\hat{x}=(1,0)$ direction. Here, we focus on cylinders with width $L_y$ and length $L_x$, where $L_y$ and $L_x$ are number of sites along the $\hat{y}$ and $\hat{x}$ directions, respectively. There are $N=L_x\times L_y$ lattice sites and the number of electrons is $N_e=N$ at half-filling, i.e., $\hat{n}_i=1$. The concentration of doped holes is defined as $\delta=\frac{N_h}{N}$ with $N_h=N-N_e$ the number of holes measured from half-filling.

For the present study, we focus on $L_y=6$ cylinders of length up to $L_x=48$ at doping concentration $\delta=1/18$, $1/12$, $1/9$ and $1/8$. We set $t=1$ as an energy unit and report results for $-0.4\leq t^\prime\leq 0.5$ with $U=12$. We perform up to 85 sweeps and keep up to $m=50,000$ states in each DMRG block with a typical truncation error $\epsilon\lesssim 3\times 10^{-6}$. For some special sets of parameters, e.g. $t'=-0.4$ and $\delta=1/18$, we have further performed the DMRG calculation with $SU(2)$ spin rotational symmetry 
by keeping an even larger $U(1)$-equivalent number of states $m\approx 100,000$ to improve the reliability and accuracy of various correlation functions at long distances. Further details of the numerical simulation are provided in the Supplemental Materials (SM).

{\bf $D$-wave SC phase: }%
The SC phase with normal $d$-wave symmetry occupies a large portion of phase diagram in the electron-doped side with positive $t'$. Similar with previous DMRG studies of doped Hubbard \cite{Jiang2020Hub} and $t$-$J$ models \cite{Jiang2018tJ,Jiang2020tJ,Gong2021,Jiang2021White,Jiang2021tJ,Jiang2022White,Jiang2023,Lu2023Sign}, this SC phase has coexisting quasi-long-range SC and charge stripe orders, but short-range spin-spin and single-particle correlations. In the following, we consider a characteristic set of parameters that is deep inside the SC phase, e.g., $t'=0.5$ and $\delta=1/12$, as an example to describe its physical properties in details.

\textit{Superconducting correlations -- } %
We first calculate the equal-time spin-singlet pair-pair correlations to find out the nature of the SC correlations defined as
\begin{eqnarray}
    \Phi_{\alpha \beta}(r)=\left<\hat{\Delta}_{\alpha}^{\dagger}(x_0,y_0) \hat{\Delta}_{\beta}(x_0+r,y_0)\right>.
\end{eqnarray}
Here $\hat{\Delta}_{\alpha}^{\dagger}(x,y) = \frac{1}{\sqrt{2}} (c^{\dagger}_{\uparrow,(x,y)} c^{\dagger}_{\downarrow,(x,y)+\alpha} - c^{\dagger}_{\downarrow,(x,y)} c^{\dagger}_{\uparrow,(x,y)+\alpha})$ is a spin-singlet pair creation operator on bond $\alpha=\hat{x}$ or $\hat{y}$. $(x_0, y_0)$ is the reference bond taken as $x_0\sim L_x/4$ and $r$ is the distance between two bonds in the $\hat{x}$ direction.

According to the Mermin–Wagner theorem, a SC state that can be realized in quasi-one-dimensional (1D) systems such as long cylinders has quasi-long-range SC correlations which decay as a power-law with the appropriate Luttinger exponent $K_{sc}$ defined by%
\begin{eqnarray}\label{Eq:Ksc}
    \Phi(r) \propto r^{-K_{sc}}. 
\end{eqnarray}
As shown in Fig.\ref{Fig:phase_SC}(a), it is clear that the spatial decay of $\Phi_{\alpha\beta}(r)$, e.g., $\Phi_{yy}(r)$, is consistent with such a power-law decay. The exponent $K_{sc}$, which is obtained by fitting the results using Eq.(\ref{Eq:Ksc}), is nearly independent of the length $L_x$ of cylinders that we have considered. For instance, the extracted $K_{sc}=1.65(3)$ and $K_{sc}= 1.66(7)$ for $\delta=1/12$ on six-leg cylinders of length $L_x=24$ and $L_x=32$, respectively. This establishes that the lightly doped Hubbard model on six-leg cylinders with positive $t'$ has quasi-long-range SC correlations as the corresponding SC susceptibility $\chi_{sc} \sim T^{- (2 - K_{sc}) }$ with $K_{sc}<2$ diverges as the temperature $T\rightarrow 0$. In addition to $\Phi_{yy}(r)$, we have also measured other components of $\Phi_{\alpha\beta}(r)$ and find that $\Phi_{yy}(r)\sim \Phi_{xx}(r) \sim - \Phi_{xy}(r)$. Contrary to the plaquette $d$-wave, the SC correlation $\Phi_{yy}$ along $\hat{y}$ direction does not change sign. Therefore, our results suggest that the SC correlations have a normal $d$-wave form. More results are provided in SM.

\textit{Charge density wave order -- } %
Similar to a previous study on four-leg cylinders \cite{Jiang2020Hub}, we have also observed a tendency to form charge stripes in the lightly doped Hubbard model on six-leg cylinders with positive $t'$. To measure the CDW order, we define the charge density $n(x,y)=\langle \hat{n}(x,y)\rangle$ and its rung average $n(x)={L_y}^{-1}\sum_{y=1}^{L_y}n(x,y)$. Fig.\ref{Fig:phase_SC}(b) shows examples of $n(x)$ on six-leg cylinders at $\delta=1/12$ with $t'=0.5$, where $x$ is the distance from one end of the cylinder up to a maximum value $x=L_x/2$. The charge density oscillations have a period $\lambda_c=\frac{1}{3\delta}$ that is consistent with ``one third-filled" charge stripes. This corresponds to an ordering wavevector $Q=6\pi\delta$ with two holes per 1D unit cell. Although this is different from the ``half-filled" charge stripes with $\lambda_c=\frac{1}{2\delta}$ on four-leg cylinders \cite{Jiang2020Hub}, it is the same with that observed in the $t$-$J$ model on six-leg cylinders with positive $t'$ \cite{Gong2021,Jiang2021tJ,Jiang2021White,Lu2023Sign}.

Similar to SC correlations, the spatial decay of CDW correlations at long distance is dominated by a power law with the Lutting exponent $K_c$ that can be obtained by fitting the charge density oscillations (Friedel oscillations) induced by the boundaries of cylinder \cite{White2002}%
\begin{eqnarray}\label{Eq:Kc}
    n(x)=A_Q \cos(Q x +\phi) x^{-K_c/2} + n_0.
\end{eqnarray}
Here $A_Q$ is an amplitude, $\phi$ is a phase shift, $n_0=1-\delta$ is the mean density and $Q=6\pi\delta$. For the characteristic set of parameters shown in Fig.\ref{Fig:phase_SC}(b), the extracted $K_c=1.5(2)$ and $K_c=1.5(1)$ for cylinders of length $L_x=24$ and $L_x=32$, respectively. The exponent $K_c$ can also be extracted independently from the charge density-density fluctuation correlation function defined by%
\begin{eqnarray}\label{Eq:NN}
    D(r)=\left<(\hat{n}_{x_0,y_0} - n_{x_0,y_0} )(\hat{n}_{x_0+r,y_0} - n_{x_0+r,y_0})\right>.
\end{eqnarray}
Here ($x_0,y_0$) is a reference site and $r$ is the distance between two sites in the $\hat{x}$ direction and $x_0\sim L/4$. Fig.\ref{Fig:phase_SC}(c) shows examples of $D(r)$ for the same parameters, the extracted value of $K_c$ using $D(r)\sim r^{-K_c}$ gives $K_c\approx 1.44$ and $K_c\approx 1.45$ for cylinders of length $L_x=24$ and $L_x=32$, respectively. Both cases are qualitatively consistent with each other as $K_c<2$ for both cases.

\textit{Spin-spin correlations -- } %
To describe the magnetic properties of the ground state, we have also calculated the spin-spin correlation function defined as%
\begin{eqnarray}\label{Eq:SS}
    F(r)=\langle \vec{S}_{x_0,y_0}\cdot \vec{S}_{x_0+r,y_0}\rangle.
\end{eqnarray}
Similar to previous studies of Hubbard model on four-leg cylinders \cite{Jiang2020Hub} and $t$-$J$ model on six-leg cylinders \cite{Jiang2021tJ,Jiang2021White,Gong2021,Lu2023Sign}, we find that $F(r)$ for finite dopings are short-ranged with a finite correlation length $\xi_s$. For example, the extracted $\xi_s$ from Fig.\ref{Fig:phase_SC}(d) using $F(r)\sim e^{-r/\xi_s}$ is $\xi_s\approx 4$. Another similar feature is that the period of $F(r)$ is characterized by a simple two-sublattice periodicity which is independent of $\delta$.

%properties of CDW 2/3 phase
\begin{figure}[bt]
\centering
\includegraphics[width=\linewidth]{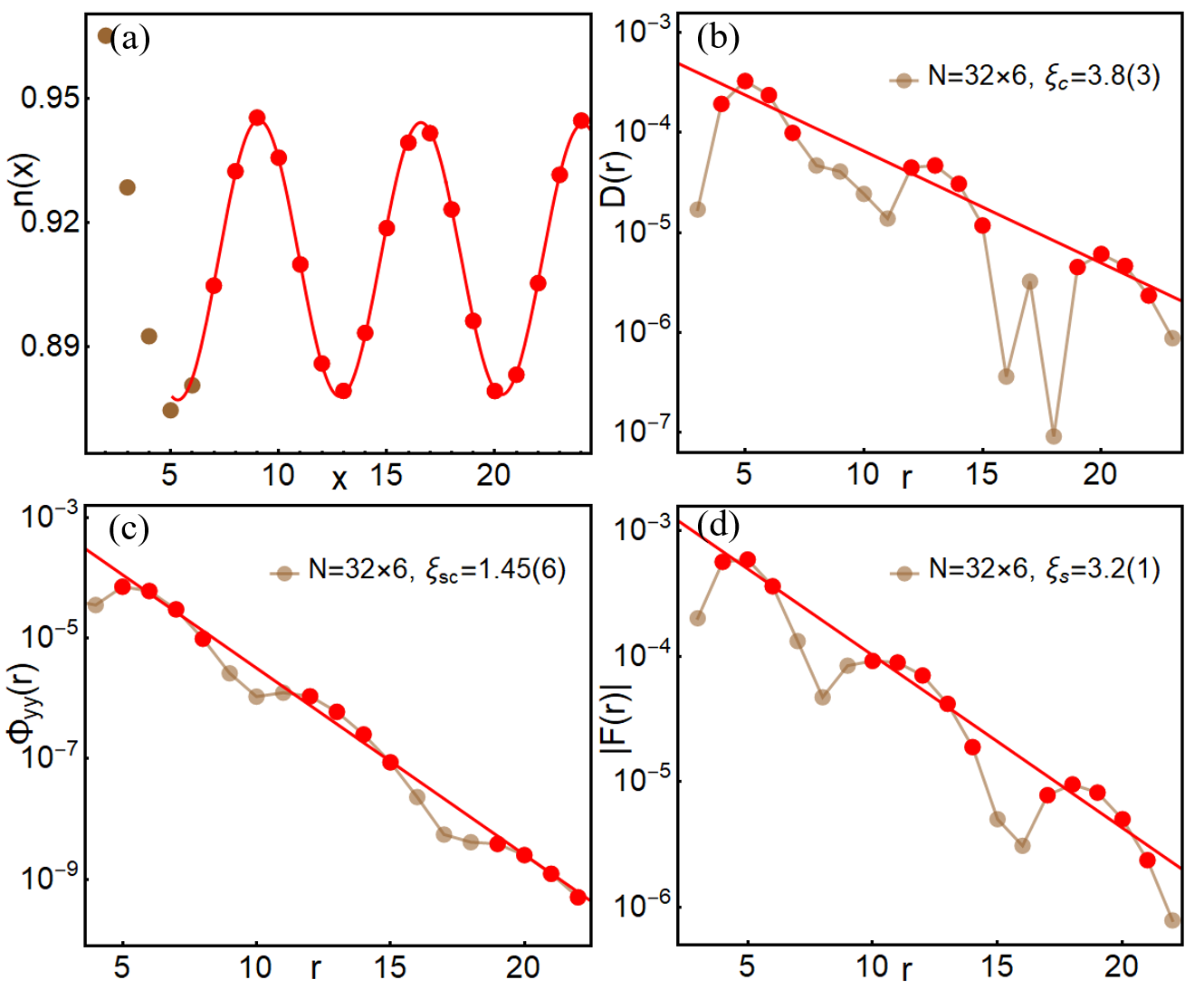}
\caption{(Color online) %Physical properties 
Characterization of the CDW-2/3 phase: (a) Charge density profile $n(x)$ fitted by Friedel oscillations (solid line) using Eq.(\ref{Eq:Kc}). 
(b) Charge density-density correlation $D(r)$ on a semi-logarithmic scale where the solid line denotes an exponential fit $D(r)\sim e^{-r/\xi_{c}}$. 
(c) Superconducting correlation $\Phi_{yy}(r)$ on a semi-logarithmic scale where the solid line denotes an exponential fit $\Phi(r)\sim e^{-r/\xi_{sc}}$. 
(d) Spin-spin correlation $F(r)$ on a semi-logarithmic scale where the solid line denotes an exponential fit $|F(r)|\sim e^{-r/\xi_s}$. Here $t'=0$, $\delta=1/12$, and data points in red are used for the fit. 
}
\label{Fig:phase_CDW2_3}
\end{figure}

{\bf CDW-2/3 phase: }%
Besides the $d$-wave SC phase, we find two distinct insulating CDW phases in Fig.\ref{Fig:PhaseDiagram}. The first CDW phase, which was referred to as ``CDW-2/3" in the blue region of phase diagram, has strong unidirectional charge stripe order but short-range SC correlations. This is similar to previous DMRG studies of the Hubbard model with $t'=0$ \cite{Jiang2019Hub,Jiang2020Hub,Ehlers2017,Qin2020}, where the system has long-range charge stripe order but short-range SC correlations. In the following, we will consider one representative set of parameters with $t'=0$ and $U=12$ at $\delta=1/2$ as an example to demonstrate the physical properties of this phase. Fig.\ref{Fig:phase_CDW2_3}(a) shows $n(x)$ with a period $\lambda_c=\frac{2}{3\delta}$ and ordering wavevector $Q=3\pi\delta$. However, in stark contrast to the $d$-wave SC phase in Fig.\ref{Fig:phase_SC}(b), the oscillation of $n(x)$ in the ``CDW-2/3" phase remains very robust whose spatial decay is nearly invisible for these lengths of ladders. Indeed, this is directly supported by our results where we find that the value of extracted $K_c=0.06(4)$ is very close to $K_c=0$. This is consistent with long-range charge stripe order. This is also supported by the short-range charge density fluctuation correlations $D(r)\sim e^{-r/\xi_c}$ having a short correlation length $\xi_c\approx 4$ in Fig.\ref{Fig:phase_CDW2_3}(b), as the critical charge fluctuation is absent in a long-range charge ordered phase.

Surprisingly in comparison with 4-leg ladders, we find that other correlations are short-ranged. For example, SC correlations, as shown in Fig.\ref{Fig:phase_CDW2_3}(c), decay exponentially as $\Phi_{yy}(r)\sim e^{-r/{\xi_{sc}}}$ with a fairly short correlation length $\xi_{sc} \approx 1.5$ lattice spacings. Similarly, spin-spin correlations decay also exponentially as $F(r)\sim e^{-r/\xi_s}$ with a short correlation length $\xi_s\approx 3.2$ close to that in the $d$-wave SC phase in Fig.\ref{Fig:phase_SC}(d). However, contrary to the $d$-wave SC phase, our results show that the spin stripes appear in this charge order phase, which are mutually commensurate with the charge stripes but with twice the wavelength.

\begin{figure}[bt]
\centering
\includegraphics[width=\linewidth]{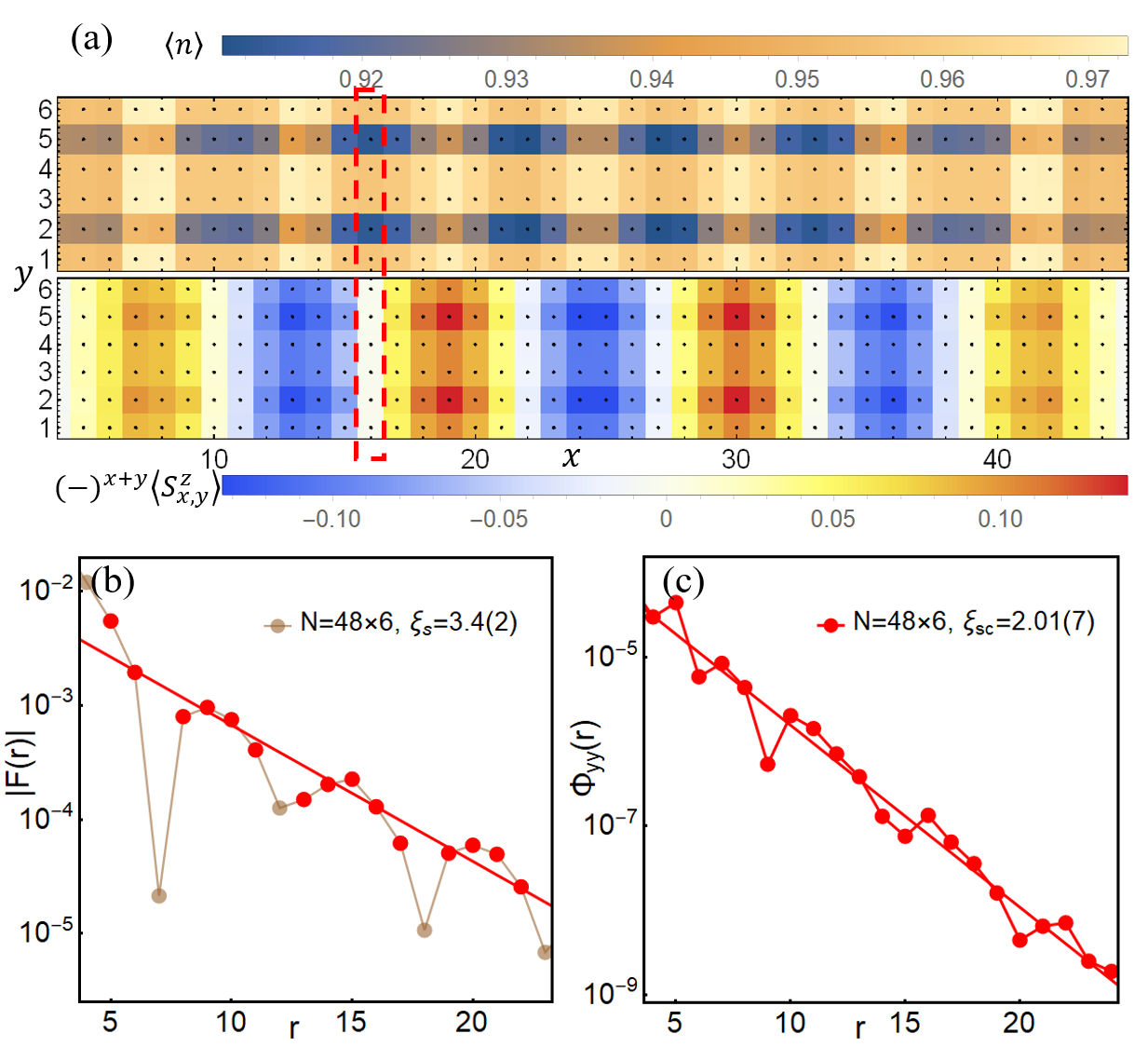}
\caption{(Color online) 
Characterization of the WC$^*$ phase: (a) Charge density profile $n(x,y)$ (upper panel) and spin density profile $S^z_{x,y}$ (lower panel). (b) Spin-spin correlation $F(r)$ on a semi-logarithm scale where the solid line denotes an exponential fit $|F(r)| \sim e^{-r/\xi_s}$. (c) Superconducting correlation $\Phi_{yy}(r)$ on a semi-logarithmic scale where the solid line denotes an exponential fit $\Phi(r)\sim e^{-r/\xi_{sc}}$. Here $t'=-0.4$, $\delta=1/18$, and data points in red are used for the fit.}\label{Fig:Wigner}
\end{figure}

{\bf Holon Wigner Crystal: }%
A second CDW phase appears in the left-bottom corner of the phase diagram in Fig.\ref{Fig:PhaseDiagram} for more negative $t'$. This is similar with the holon Wigner crystal, which was referred to as WC$^*$, reported in previous DMRG studies in doped spin liquids on the Kagome lattice \cite{Jiang2017,Peng2021}. Distinct with the ``CDW-2/3" phase, the CDW order (see Fig.\ref{Fig:Wigner}(a)) in the WC$^*$ phase breaks translational symmetries along both the $\hat{x}$ and $\hat{y}$ directions \cite{Jiang2021White}. The entire charge density profile appears to prefer a rectangular lattice with an emergent larger unit cell each containing one of the blue spots. The number of these emergent unit cells is equal to the number of doped holes. Therefore, this is not a crystal of hole pairs.

Interestingly, similar with previous studies of doped Kagome lattice spin liquids \cite{Jiang2017,Peng2021}, our results suggest that this CDW state is consistent with a Wigner crystal of holons, where the holon carries the charge of an electron but without its spin\cite{Rokhsar1988}, rather than doped holes. This is because if this is a Wigner crystal of doped holes, the spin and charge degrees of freedom of doped holes will be bound together without spin-charge separation. However, this is inconsistent with our results.
In the lower panel of Fig.\ref{Fig:Wigner}(a), we show the modified spin density profile $(-1)^{x+y}\langle S^z_{x,y}\rangle$ on a six-leg cylinder of length $L_x=48$ by keeping $m=40,000$ number of states, where a clear spin-charge separation is observed. Firstly, the spin density profile $(-1)^{x+y}\langle S^z_{x,y}\rangle$ exhibits clear anti-phase domain-walls only in the $x$ direction, which is distinct from the broken translation symmetries along both $x$ and $y$ directions in the charge density profile. Secondly, the maxima of $|\left< S^z_{x,y} \right>|$, i.e., the red and blue spots in the lower panel in Fig.\ref{Fig:Wigner}(a), appear exactly at the minima of the hole density profile, i.e., $1-\langle n(x,y)\rangle$, rather than its maxima. Both of these suggest the presence of spin-charge separation which is inconsistent with a Wigner crystal formed by doped holes. Similar charge and spin density profiles are observed in lightly doped $U=8$ Hubbard models with $t'=-0.4$. (see SM for details.)

A more direct evidence to support the Wigner crystal of holons is the existence of gapped spin excitations evidenced by exponentially decaying spin-spin correlations. To remove a residual $\left< S^z(x,y) \right>$ that breaks $SU(2)$ spin rotational symmetry, which is retained in the simulation even when keeping $m=40,000$ states, we set the reference point ($x_0,y_0$) of $F(r)$ to the location of a holon, e.g. $(x_0,y_0)=(16,2)$ for the six-leg cylinder of length $L_x=48$ in Fig.\ref{Fig:Wigner}(a), and further improve the accuracy of our simulation by performing other DMRG calculations having $SU(2)$ symmetry \cite{McCulloch2002} and keeping up to $m=31,000$ $SU(2)$ states (effectively $m\sim 100,000$ $U(1)$ states). Consistent with that of a WC$^*$, we find that spin-spin correlations decay exponentially as $F(r)\sim e^{-r/\xi_s}$ as shown in Fig.\ref{Fig:Wigner}(b). The spin-spin correlation length is $\xi_s\approx 3$ which is shorter than the separation between two adjacent holons in the $\hat{x}$ direction. This is a clear signature of spin-charge separation which is consistent with a holon Wigner crystal.

Similar to the ``CDW-2/3" phase, SC correlations also decay exponentially as $\Phi(r) \sim e^{-r/\xi_{sc}}$ with a short correlation length $\xi_{sc}\approx 2$. The pairing symmetry is consistent with that of a $d$-wave form evidenced by $\Phi_{yy}(r)\sim -\Phi_{xy}(r)\sim \Phi_{xx}(r)$. This is qualitatively distinct from the quasi-long-range SC with ``plaquette" $d$-wave symmetry on four-leg square cylinders with negative $t'$  \cite{Jiang2017,Jiang2019Hub,Jiang2020Hub,Chung2020,Peng2022}.

{\bf Summary and discussion: }% 
We have studied the ground state properties of the lightly doped $t$-$t'$-$U$ Hubbard model on six-leg square cylinders, and find that its phase diagram is very sensitive to both $t'$ and $\delta$. In the electron doped case with positive $t'$, we find a robust $d$-wave SC phase with coexisting quasi-long-range SC and CDW orders. This SC phase shares many similarities with the one previously reported for doped Hubbard model on four-leg cylinders \cite{Jiang2020Hub}, and the $t$-$J$ model on both six and eight-leg cylinders \cite{Gong2021,Jiang2021White,Jiang2021tJ,Jiang2022White,Jiang2023,Lu2023Sign}. Therefore, our results suggest that long-range SC could also be realized in the electron-doped Hubbard model in two dimensions. In the hole-doped case with negative $t'$, our results show that the Hubbard model on six-leg cylinders is not SC but possesses long-range CDW order. Although this is in stark contrast to four-leg cylinders \cite{Jiang2019Hub,Jiang2020Hub,Chung2020,Peng2022} with quasi-long-range superconductivity, it is consistent with hole-doped $t$-$J$ model, i.e., the strong coupling limit of the Hubbard model, on both six and eight-leg square cylinders with negative $t'$ \cite{Gong2021,Jiang2021White,Jiang2021tJ,Jiang2022White,Jiang2023,Lu2023Sign}.

In the vicinity of $t'=0$ line, a charge stripe phase, which is similar with the one reported in recent DMRG studies in the ``pure" Hubbard model \cite{Ehlers2017,Jiang2019Hub,Jiang2020Hub,Qin2020}, occupies a fairly large portion of the phase diagram. However, it appears to be very sensitive to positive $t'$ where a small $t'$ can drive the system into a phase separation (PS) close to $\delta=1/8$ in an extended region of $t'$. It will be interesting to study whether such a PS can be suppressed by introducing additional terms such as further-neighbor electron hopping and extended electron interaction so that a SC phase can be realized accordingly. Answering these questions may lead to better understanding of
the mechanism of high temperature superconductivity.

Quite surprisingly, at relatively lower doping level and larger negative $t'$, we find a novel insulating CDW phase, i.e., WC$^*$, which is similar with the one reported for doped Kagome lattice spin liquids \cite{Jiang2017,Peng2021} and square lattice $t$-$J$ model \cite{Jiang2021White}. This is a Wigner crystal of holons, rather than either doped holes or hole pairs. Given that the fractional excitation is unlikely to show up in the undoped Hubbard and Heisenberg models (with corresponding $J_2/J_1\sim 0.1$), this novel WC$^*$ phase, which appears to be a doping induced fractional phase, is quite striking. It will be interesting to understand its microscopic origin which we will leave in a future study.

\emph{Acknowledgments:} We would like to thank Steven Kivelson, Jan Zaanen, Zheng-Yu Weng and Shengtao Jiang for insightful discussions and suggestions. Y.-F.J. acknowledges support from the National Program on Key Research Project under Grant No.2022YFA1402703 and Shanghai Pujiang Program under Grant No.21PJ1410300. T.P.D. and H.C.J. are supported by the Department of Energy (DOE), Office of Sciences, Basic Energy Sciences, Materials Sciences and Engineering Division, under Contract No. DE-AC02-76SF00515.

\textit{Note added -- }%
We have become aware of an independent but closely related study of doped $t$-$t'$-$U$ Hubbard model on square lattice using a combination of DMRG and constrained path auxiliary field quantum Monte Carlo \cite{Xu2023}, which reports results on quasi-two-dimensional square lattice by applying a finite global $d$-wave pairing field. Conversely, our study focuses on long six-leg cylinders at $U$=12 without applying pinning fields.

%==Reference==
%\bibliography{Refs}
%\clearpage
\clearpage

\begin{widetext}

\renewcommand{\theequation}{S\arabic{equation}}
\setcounter{equation}{0}
\renewcommand{\thefigure}{S\arabic{figure}}
\setcounter{figure}{0}
\renewcommand{\thetable}{S\arabic{table}}
\setcounter{table}{0}

%==appendix==
\section{Supplemental Material}
\subsection{Superconducting pair-pair correlations in the $d$-wave SC phase}
In Fig.\ref{FigSM:SC}, we show more results for SC correlations deep inside the $d$-wave SC phase with $t'=0.5$ at $\delta=1/12$. The pairing symmetry of SC correlations $\Phi_{\alpha\beta}$ is consistent with 2D $d$-wave as demonstrated in Fig.\ref{FigSM:SC}(a) with $\Phi_{xx}(r) \sim \Phi_{yy}(r)\sim -\Phi_{xy}(r)$. In Fig.\ref{FigSM:SC}(b) and (c), we provide examples showing how we perform a finite-truncation-error extrapolation for SC correlations for a zero truncation-error limit $\epsilon \rightarrow 0$ (i.e., $m\rightarrow \infty$ ). Taking $\Phi_{yy}(r)$ as an example, we first calculate $\Phi_{yy}(r)$ by keeping $m = 20000 \sim 45000$ states as shown in Fig.\ref{FigSM:SC}(b). Then an extrapolation using a second order polynomial $\Phi(\epsilon) = \Phi_0 + a_1 \epsilon + a_2 \epsilon^2$ is used to extract $\Phi_{yy}(r)$ for each $r$ to the zero truncation-error limit as shown in Fig.\ref{FigSM:SC}(c). Here $\epsilon$ is the truncation error associated with the number of states $m$, $a_1$ and $a_2$ are fitting parameters. 

\begin{figure}[h]
\centering
\includegraphics[width=\linewidth]{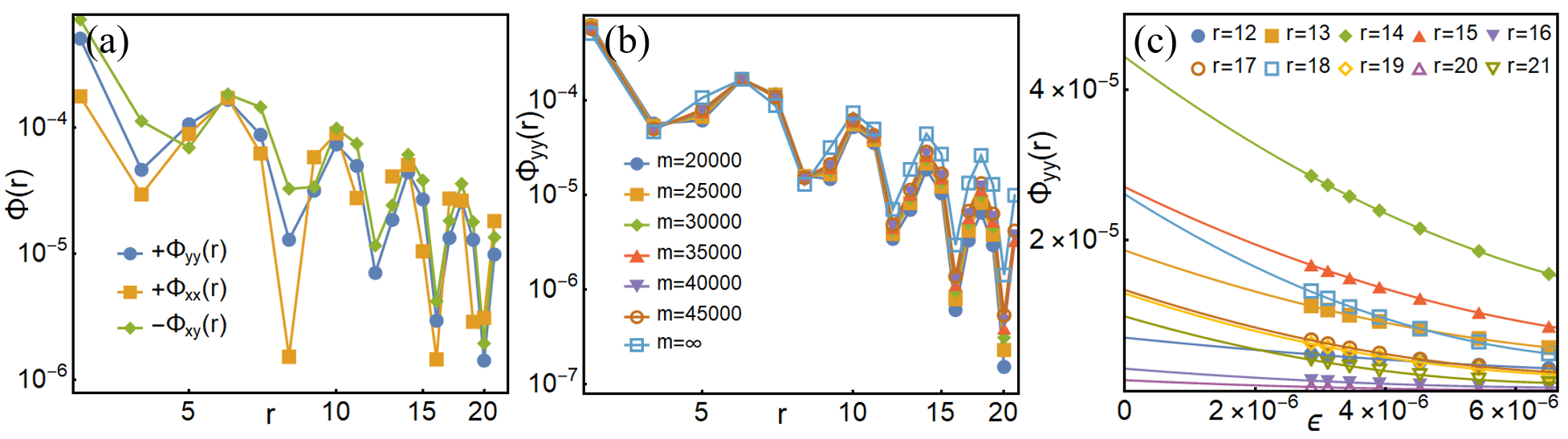}
\caption{(Color online) Convergence of superconducting correlations in the $d$-wave SC phase on $N=32\times 6$ cylinder. 
(a) $\Phi_{xx}(r)$, $\Phi_{xy}(r)$ and $\Phi_{yy}(r)$. 
(b) $\Phi_{yy}(r)$ by keeping $m=20000 \sim 45000$ states and its extrapolation in the limit $m\rightarrow \infty$. 
(c) Examples of finite truncation error extrapolation of $\Phi_{yy}(r)$ using a second order polynomial (solid lines) for $r=12\sim 21$. 
Here $t'=0.5$ and $\delta=1/12$.}
\label{FigSM:SC}
\end{figure}

\subsection{Results for the holon Wigner crystal phase}
\begin{figure}[h]
\centering
\includegraphics[width=0.7\linewidth]{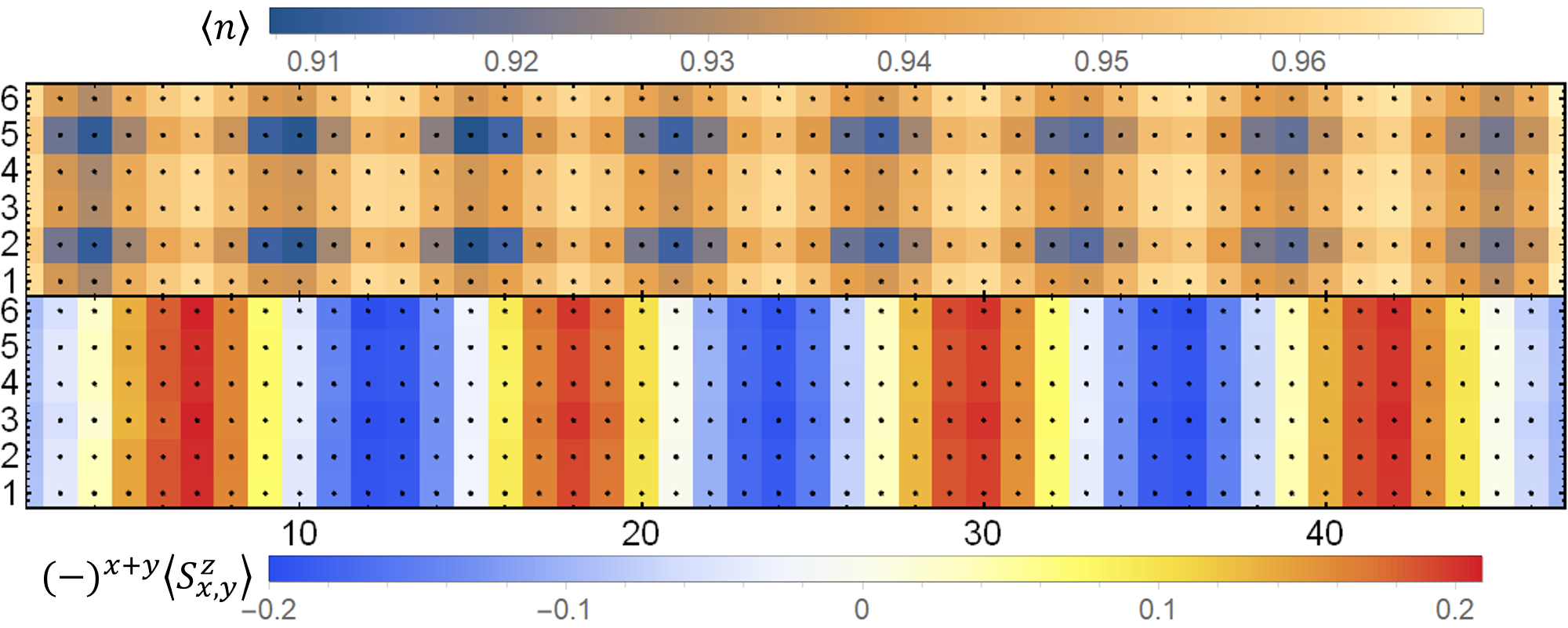}
\caption{(Color online) Charge density profile $n(x,y)$ (upper panel) and spin density profile $S^z_{x,y}$ (lower panel) in the WC$^*$ phase on $N=48\times 6$ cylinder with $U=8$ and $t'=-0.4$ at $\delta=1/18$ obtained by keeping $m=16000$ states. The number of emergent unit cells (blue spots) is equal to the number of doped holes.}
\label{FigSM:hWC}
\end{figure}

We provide more examples to show that the WC$^*$ phase is also stable for the $U=8$ Hubbard model with negative $t'$. Fig.\ref{FigSM:hWC} shows the density profile $n(x,y)$ and modified spin density profile $(-1)^{x+y}\left<S^z_{x,y}\right>$ of the $U=8$ doped Hubbard model with $t'=-0.4$ at $\delta=1/18$ on $N=48\times 6$ cylinder. It is clear that $n(x,y)$ breakstranslation symmetries along both $\hat{x}$ and $\hat{y}$ directions, and the number of the emergent CDW unit cells is equal to the number of doped holes. The spin density profile shows a clear signature of spin-charge separation as discussed in the main text. Similar density profiles have been observed in the WC$^*$ phase including the sets of parameters $(t', \delta)=(-0.4, 1/12)$, $(-0.5, 1/18)$ and $(-0.25, 1/18)$.

\end{widetext}

\end{document}